\documentclass[aps,prb,twocolumn,a4,10pt,floatfix,superscriptaddress]{revtex4-1}
\pdfoutput=1
\usepackage{verbatim}
\usepackage{amsmath,amssymb,amsfonts,upgreek}
\usepackage{graphicx}
\usepackage{setspace}
\usepackage{bm}
\usepackage{multirow}
\usepackage{hyperref}

\usepackage{color}
\usepackage{tabularx}
\usepackage{booktabs}

\newcommand{\btfo}{\ensuremath{\mathrm{Bi_5FeTi_3O_{15}}}}
\newcommand{\bio}{\ensuremath{\mathrm{Bi_2O_2}}}

\newcommand{\fe}{\ensuremath{\mathrm{Fe^{3+}}}}
\newcommand{\ti}{\ensuremath{\mathrm{Ti^{4+}}}}

\begin{document}

\title{Magnetic order in 4-layered Aurivillus phases}

\author{Axiel Ya\"el Birenbaum} 
\email{birenbaumyl@ornl.gov} 
\altaffiliation[\\Present address: ]{Materials Science and Technology
  Division, Oak Ridge National Laboratory, 1 Bethel Valley Rd., Oak
  Ridge, TN, USA}
\affiliation{Materials Theory, ETH Z\"urich, Wolfgang-Pauli-Strasse
  27, 8093 Z\"urich, Switzerland}
\author{Andrea Scaramucci} 
\email{andrea.scaramucci@mat.ethz.ch}
\affiliation{Materials Theory, ETH Z\"urich, Wolfgang-Pauli-Strasse
  27, 8093 Z\"urich, Switzerland} 
\affiliation{Laboratory for Scientific Development and Novel Materials, Paul Scherrer
  Institut, CH-5232 Villigen PSI, Switzerland}
\author{Claude Ederer} 
\email{claude.ederer@mat.ethz.ch} 
\affiliation{Materials Theory, ETH Z\"urich, Wolfgang-Pauli-Strasse
  27, 8093 Z\"urich, Switzerland}

\date{\today}

\begin{abstract}
We determine the viability of 4-layered Aurivillius phases to exhibit
long-range magnetic order above room temperature. We use Monte Carlo
simulations to calculate transition temperatures for an effective
Heisenberg model containing a minimal set of required couplings. The
magnitude of the corresponding coupling constants has been determined
previously from electronic structure calculations for \btfo, for which
we obtain a transition temperature far below room temperature. We
analyze the role of further neighbor interactions within our Heisenberg
model, in particular that of the second-nearest-neighbor coupling
within the perovskite-like layers of the Aurivillius structure, as
well as that of the weak inter-layer coupling, in order to identify
the main bottleneck for achieving higher magnetic transition
temperatures. Based on our findings, we show that the most promising
strategy to obtain magnetic order at higher temperatures is to
increase the concentration of magnetic cations within the
perovskite-like layers, and we propose candidate compounds where
magnetic order could be achieved above room temperature.
\end{abstract}

\maketitle

%------------------------------------

\section{Introduction} 
\label{sec:intro}

Coexistence of ferroelectricity and magnetic long-range order in the
same material is an uncommon phenomenon \cite{Spaldin2000}. Materials
that exhibit such coexistence are called magnetoelectric multiferroics
\cite{Schmid:1994}. In recent years, such materials have attracted
huge interest,\cite{Spaldin/Fiebig:2005,Spaldin/Cheong/Ramesh:2010}
motivated in large parts by the possibility of exploiting the
coexistence of the two types of long-range order to create four-state
logic devices,\cite{Gajek_et_al:2007} and by the prospect of using
cross-couplings to switch magnetic bits with an applied
voltage.\cite{Bibes/Barthelemy:2008} Two main routes that can lead to
magnetoelectric multiferroic states are often distinguished
\cite{Khomskii2009}. The first one requires the presence of particular
types of magnetic order that break inversion symmetry (e.g. spiral
states in $R$MnO$_3$ \cite{Cheong/Mostovoy:2007}) and hence can couple
to electric polarization. The second route is realized in compounds in
which ferroelectricity is induced by a structural instability that is
only weakly affected by magnetic order. A prominent example for this
second scenario is BiFeO$_3$, which is one of the most extensively
studied multiferroics.\cite{Catalan/Scott:2009} Due to the relative
scarcity of magnetoelectric multiferroics, the design of new materials
with robust multiferroic properties above room temperature is very
desirable.

\begin{figure}
\centerline{\includegraphics[width=0.9\columnwidth]{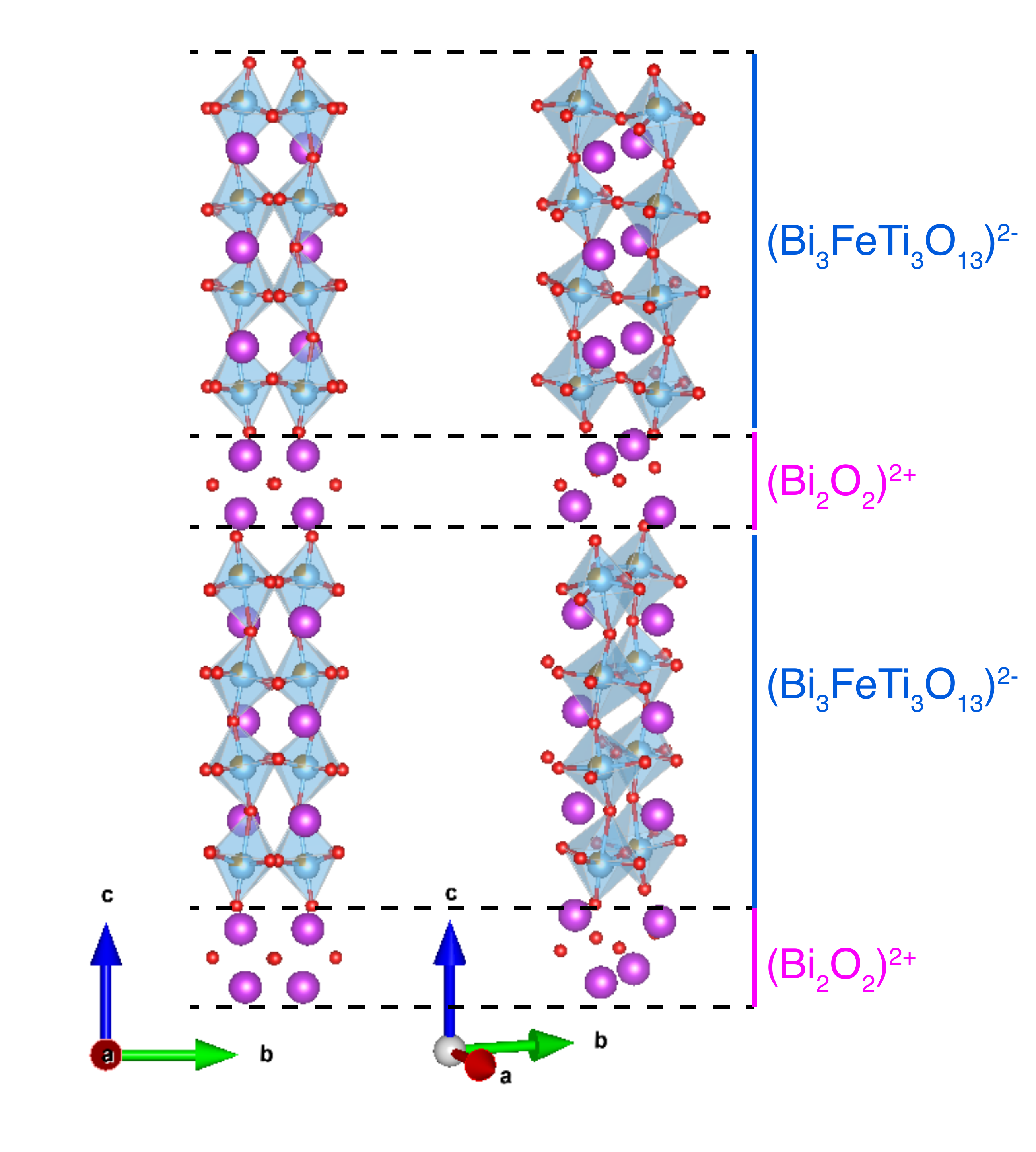}}
\caption{(Color online) Structure of \btfo, an Aurivillius phase
  corresponding to $m=4$, at temperatures below the ferroelectric
  transition, viewed from two slightly different perspectives. Purple
  and red spheres indicate Bi$^{3+}$ and O$^{2-}$ ions,
  respectively. The ions inside the blue octahedra ($B$ sites) are
  Ti$^{4+}$ and Fe$^{3+}$ with concentrations $x(\text{Ti})=3/4$ and
  $x(\text{Fe})=1/4$.}
\label{fig:Structure}
\end{figure}

The Aurivillius phases are a family of naturally-layered oxides.
Their structure consists of $m$ perovskite units,
($A_{m-1}B_m$O$_{3m+1}$)$^{2-}$, stacked along the $c$ direction, and
alternating with fluorite-like layers of (Bi$_2$O$_2$)$^{2+}$ (see
Fig.~\ref{fig:Structure} for a case with
$m=4$).~\cite{Newnham/Wolfe/Dorrian:1971} Aurivillius phases are well
known for their good ferroelectric properties with Curie temperatures
well above room temperature and low
fatigue\cite{PazdeAraujo_et_al:1995,Park_et_al:1999}.
Furthermore, Aurivillius phases display a large chemical flexibility,
which allows to incorporate various magnetic (and non-magnetic)
cations on the octahedrally coordinated $B$ sites, making this class
of materials a very promising starting point for engineering high
temperature multiferroics.  In particular, Aurivillius systems
incorporating $3d$ transition metal cations, which generally can give
rise to high N{\'e}el temperatures in oxides, are natural candidates
to explore new multiferroic compounds.

Experimentally, most efforts so far have focused on $m=4$ structures
with 75\,\% \ti\ and 25\,\% \fe\ (or other magnetic cations with
valence $+3$) distributed over the $B$
sites.\cite{Dong_et_al:2008,Mao_et_al:2009,Giddings_et_al:2011,Chen_et_al:2014}
However, conflicting magnetic properties have been reported for these
compounds. In the most studied case of \btfo, Srinivas \textit{et al.}
have reported an antiferromagnetic transition temperature of
80~K\cite{Srinivas:1999}, which conflicts with more recent studies
that report paramagnetic behavior with no magnetic long-range order
down to very low
temperatures.\cite{Dong_et_al:2008,Chen_et_al:2013,Jartych_et_al:2013}
Furthermore, it was shown that the observed properties often depend
strongly on the synthesis conditions and can be caused by trace
amounts of impurity phases, which are very hard to detect using
standard laboratory-based characterization methods
\cite{Keeney_et_al:2012}. Therefore, it is of great interest to have a
theoretical estimation of the expected magnitude of the temperature,
$T_C$, at which magnetic order might arise.

From the theoretical side, two factors appear crucial for the possible
emergence of long-range magnetic order in Aurivillius phases. The
first is the relatively low concentration of magnetic cations (only
25\,\% on the perovskite $B$ sites in the above examples), the second
is the short range of the superexchange interaction, which is the
dominant coupling mechanism between magnetic ions in insulating
oxides.

The concentration of magnetic cations is fixed by stoichiometric
constraints. As already mentioned, most systems that have been studied
so far correspond to 4-layered ($m=4$) Aurivillius phases with
composition Bi$_{5}M$Ti$_3$O$_{15}$, where $M$ is a trivalent $3d$
transition metal cation. Note that due to the different valence it is
not possible to simply substitute some of the Ti$^{4+}$ by $M^{3+}$ to
increase the concentration of magnetic cations on the $B$ sites. In
Sec.~\ref{sec:higher_conc} we discuss possible alternative compositions
that allow to increase the concentration of magnetic cations, but
first we will focus on already existing compounds with $m=4$ and
25\,\% magnetic cations on the $B$-sites, in particular the case with
$M$=\fe.

For this relatively low concentration of magnetic ions, the short
range of the superexchange mechanism represents a serious challenge
for obtaining long-range magnetic order at high
temperature. Typically, superexchange can be very strong between
magnetic ions sharing the same ligand (i.e. nearest-neighbor
superexchange), but decreases quickly with distance (or, more
precisely, with the number of bonds along the shortest superexchange
path connecting two magnetic ions).

The low concentration of magnetic
ions (25\,\% in the above examples) results in a low average number of
magnetic ions that share an oxygen ligand and thus are strongly
coupled. In the worst case, the percolation threshold for a fully
connected network of nearest-neighbor superexchange bonds might not be
reached, and the system simply consists of many isolated clusters that
are only weakly coupled by further-neighbor superexchange
interactions. However, even somewhat above this percolation threshold,
the effective dimensionality of the resulting ``magnetic lattice'' can
be quite low. This implies that the much weaker further-neighbor
superexchange couplings are playing a crucial role for achieving
long-range order in \btfo\ and similar Aurivillius phases. In
particular, due to the presence of the (Bi$_2$O$_2$)$^{2+}$ layers
(see Fig.~\ref{fig:Structure}), there are no strong nearest-neighbor
links between adjacent perovskite blocks, which are only coupled
through rather weak further neighbor superexchange (the shortest
possible superexchange path between two adjacent perovskite blocks is
along a sequence of $M$-O-O-O-$M$ bonds). However, coupling across the
(\bio)$^{2+}$ layers is essential to achieve long-range order along
the $c$ direction, and its small size might critically affect the
transition temperature.

For the case of \btfo, magnetic coupling constants have been
calculated using first principles electronic structure
calculations.~\cite{Birenbaum:2014bl} Indeed, a rather strong coupling
for \fe\ cations in nearest-neighbor positions of around 45\,meV,
corresponding to a temperature scale of $\sim$ 520\,K has been
obtained. In contrast, the coupling between \fe\ cations in
next-nearest-neighbor positions is more than one order of magnitude
smaller (1-2\,meV, corresponding to $\sim$ 15\,K), and the inter-layer
coupling was estimated to be around 0.3\,meV ($\sim$ 3.5\,K).

In this article, we present Monte Carlo simulations for an effective
Heisenberg model applicable to $m=4$ Aurivillius systems. Based on the
coupling strengths calculated for \btfo, we obtain an upper bound for
the magnetic transition temperature ($T_\text{C}$) of this material of
22\,K, i.e. significantly below room temperature.
In order to explore which of the two weak further-neighbor couplings
represents the more severe bottleneck for obtaining high $T_\text{C}$,
we individually vary the strength of the next-nearest-neighbor
coupling within the perovskite blocks as well as that of the
inter-layer coupling. These calculations show that, for a magnetic ion
concentration of 25\,\%, the next-nearest-neighbor interaction is
crucial to achieve good percolation (and thus high $T_\text{C}$)
within the perovskite blocks, but that the influence of the weak
inter-layer coupling is less severe. We then also vary the
concentration of magnetic ions on the $B$ sites of the model, using
realistic magnitudes for the coupling constants based on the
calculated values for \btfo. We show that high transition temperatures
can in principle be achieved for concentrations above 50\,\%, in spite
of the rather weak inter-layer coupling. Finally, we propose a route
to achieve such higher concentrations of magnetic ions by substituting
\ti\ with non-magnetic cations with higher valence, such as Nb$^{5+}$,
Ta$^{5+}$, Mo$^{6+}$, or W$^{6+}$.

The paper is structured as follows. In Sec.~\ref{sec:theo_meth} we
first describe our model, discuss its limitations, and clarify which
couplings beyond nearest-neighbor we consider. In the same section we
also specify the technical details of our simulations. In
Sec.~\ref{sec:BTFO} we then present and discuss the results obtained
for \btfo\ and investigate the dependence of $T_\text{C}$ on the
strength of further-neighbor couplings within and in between the
$m$-perovskite blocks. In Sec.~\ref{sec:higher_conc} we discuss the
dependence of the transition temperature on the concentration of
magnetic ions, and we propose possible compositions with
concentrations larger than 25\,\%. Finally, in Sec.~\ref{sec:conclu}
we summarize our main conclusions.

\section{Model and Methods} 
\label{sec:theo_meth}

\subsection{Model and Couplings in 4-Layered Aurivillius Phases} 
\label{subsec:parameters}

In this section we discuss the effective model that we use for a
generic value of magnetic ion concentration, $x$, on the $B$ sites of
an $m=4$ Aurivillius phase. As the network of superexchange bonds is
determined by the specific distribution of magnetic and non-magnetic
cations over the available $B$ sites, we first discuss the method by
which we distribute the magnetic ions within our simulation
cell. Then, we describe the relevant magnetic couplings that we
consider, and finally we define the specific Heisenberg Hamiltonian
that we use in our simulations.

\begin{figure}
\centering
\includegraphics[width=\columnwidth]{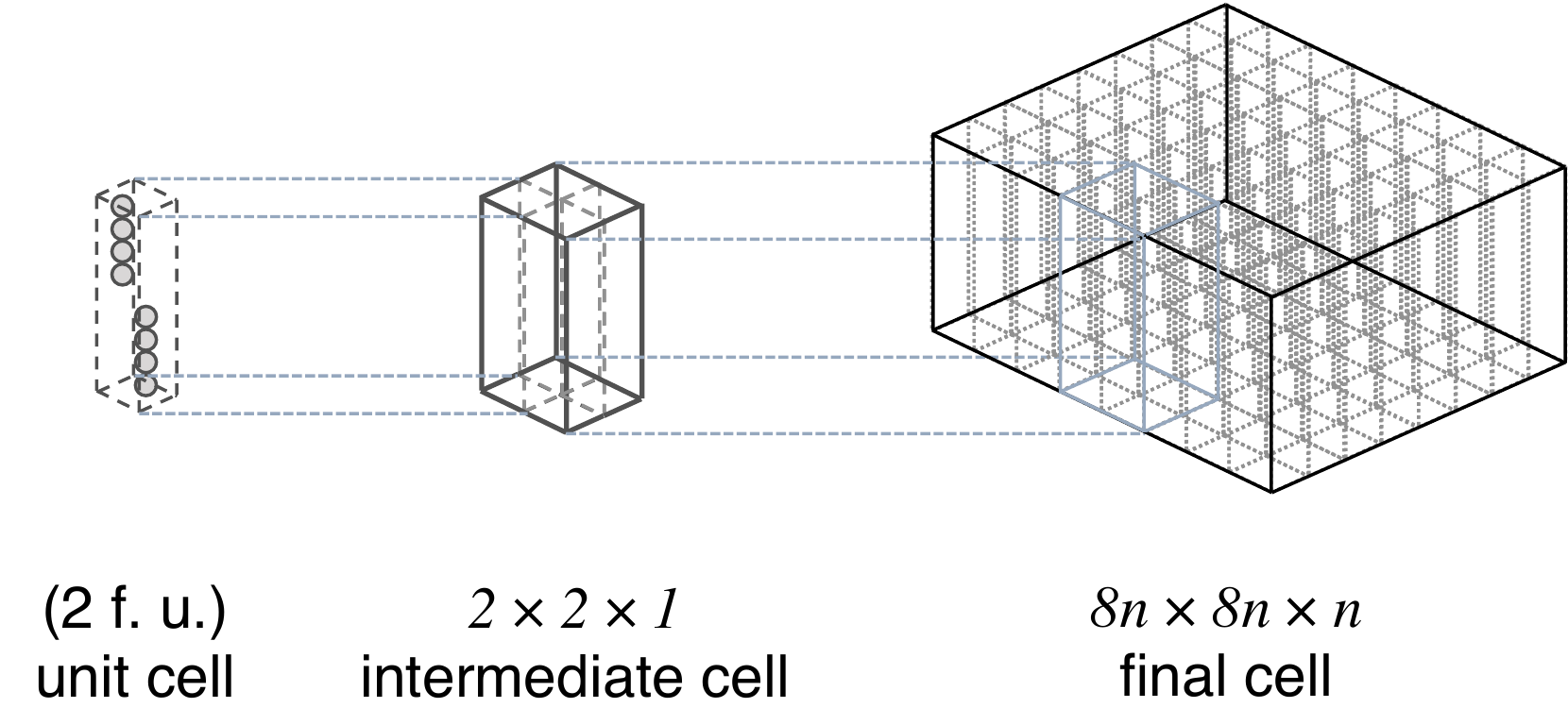}
\caption{Construction of the cell used to simulate a 4-layered
  Aurivillius phase. Our basic unit cell contains 2 formula units
  (f.u.). An intermediate cell is constructed as $2 \times 2 \times 1$
  supercell of the basic cell, and $x \cdot 32$ magnetic ions are
  randomly distributed over the 32 $B$ sites of this intermediate
  cell. The final simulation cell is then constructed by stacking
  several of such intermediate cells to form a supercell composed of
  $8n \times 8n \times n$ unit cells ($n=1$ in the figure).}
\label{fig:cell}
\end{figure}

The conventional (primitive orthorhombic) unit cell of the
ferroelectric structure of \btfo\ and related 4-layered Aurivillius
phases with $A2_1am$ space group symmetry consists of four formula
units (f.u.). However, for our model we need to consider only the
sites that can potentially be occupied by magnetic cations, and thus
the cell we use in our simulations is based on the simple tetragonal
cell sketched on the left in Fig.~\ref{fig:cell}. This cell contains
two f.u., or $2 \times 4$ $B$ sites, stacked along the $c$
direction. The lateral shift between subsequent groups of 4 $B$ sites
is related to the \bio-layer in between two 4-perovskite blocks.

We then construct an $8n \times 8n \times n$ supercell of the basic
cell. This supercell contains $64 n^3$ $B$ sites, over which we have
to distribute a total of $N = x \cdot 64n^3$ magnetic cations. In
order to avoid creating configurations with an extremely inhomogeneous
distribution of magnetic cations, we first divide the full supercell
into \emph{intermediate} cells (see middle part of
Fig.~\ref{fig:cell}), and then randomly distribute the magnetic
cations with the constraint that stoichiometry is satisfied within
each intermediate cell. Thus, the resulting concentration of magnetic
sites is exactly equal to $x$ within each intermediate cell. We note
that configurations where stoichiometry would be violated over larger
volumes will be strongly disfavored in the real material due to the
Coulomb interaction. We use a size of $2 \times 2\times 1$ in units of
the basic cell for these intermediate cells.

Note that we use a supercell with 8 times more basic unit cells along
the two in-plane directions than along the $c$ direction, in order to
obtain an aspect ratio of the final simulation cell close to one (in
terms of number of adjacent $B$ sites). This is also expected to
reduce the temperature range for which the correlation length becomes
comparable with the system size. Due to the rather weak inter-layer
coupling between the $m$-perovskite layers, the correlation length is
expected to be smaller along the $c$ direction than along the
perpendicular in-plane directions.

We point out that, apart form the stoichiometry constraint on the
intermediate cell level, we use a completely random (homogeneous)
distribution of magnetic cations over the available sites. Even though
Ref.~\onlinecite{Birenbaum:2014bl} reported a preference of \fe\ to
occupy the inner site in \btfo, this tendency is not very
strong. Furthermore, a homogeneous random distribution of magnetic
ions is presumably most favorable for the development of long-range
magnetic order, and since we primarily want to establish an upper
bound for the magnetic transition temperature, we are focusing here on
this most favorable case.
Furthermore, we make no assumptions on potential correlations between
the relative positions of magnetic ions on neighboring sites.

Once a specific distribution of magnetic ions on the $B$ sites is
constructed, it is possible to establish the network of exchange
couplings connecting the magnetic sites.
As discussed in Ref.~\onlinecite{Birenbaum:2014bl}, one can in
principle distinguish four different, symmetry-inequivalent couplings
between magnetic ions in ``nearest-neighbor'' positions within the
perovskite blocks. However, {\it ab-initio} calculations for
\btfo\ presented in Ref.~\onlinecite{Birenbaum:2014bl} also show that
at least three of these nearest-neighbor couplings have very similar
strength.
For simplicity, and since our main goal is to establish an upper bound
for the magnetic transition temperature, we consider all
nearest-neighbor couplings to be identical for the purpose of this
work, and we denote the corresponding coupling strength by
$J_{\text{NN}}$.

Now, considering only the network of spins connected by
$J_{\text{NN}}$, each of the 4-perovskite blocks is effectively a
simple cubic lattice with only four layers along $c$. Each site of
this lattice is then randomly occupied with probability $x$. The
problem of finding the minimal average occupation of sites, $x_c$,
necessary to obtain site percolation through nearest-neighbor bonds in
a simple cubic lattice has been studied
extensively,~\cite{Sykes1964,Grassberger1992,Jan1998,Deng2005,Kurzawski:2012fx}
and the critical concentration was found to be $x_c \approx 0.312$.
This implies that for the case of the 4-layered Aurivillius phases
with composition Bi$_5M$Ti$_3$O$_{15}$, i.e. $x=0.25$, no magnetic
long-range order can be obtained by considering only $J_{\text{NN}}$,
and thus further-neighbor couplings play an essential role for the
magnetic ordering.
Therefore, we also include magnetic coupling between
next-nearest-neighbor positions within the $m$-perovskite blocks into
our model. The values of some of these couplings for the case of
\btfo\ have also been calculated in Ref.~\onlinecite{Birenbaum:2014bl}
and, similar to the nearest-neighbor couplings, can be viewed as
approximately constant and independent of the specific
next-nearest-neighbor configuration. Thus, we use the same coupling
constant, $J_{\text{NNN}}$, for all next-nearest-neighbor bonds within
the $m$-perovskite blocks of our model.

Although the presence of such a next-nearest-neighbor coupling
guarantees to overcome the percolation threshold within the
$m$-perovskite blocks for a concentration of $x=0.25$ ($x_c \approx
0.137$ for a simple cubic lattice with both nearest and next-nearest
neighbor couplings~\cite{Kurzawski:2012fx}), so far our model does not
include any coupling between adjacent $m$-perovskite
blocks. Therefore, to achieve three-dimensional long-range magnetic
order it is essential to also consider a coupling between closest
neighbors across a \bio-layer, i.e., between two adjacent
$m$-perovskite blocks. We denote this inter-layer coupling by
$J_{\text{INTER}}$.

\begin{figure}
\centering
\includegraphics[width=0.75\columnwidth]{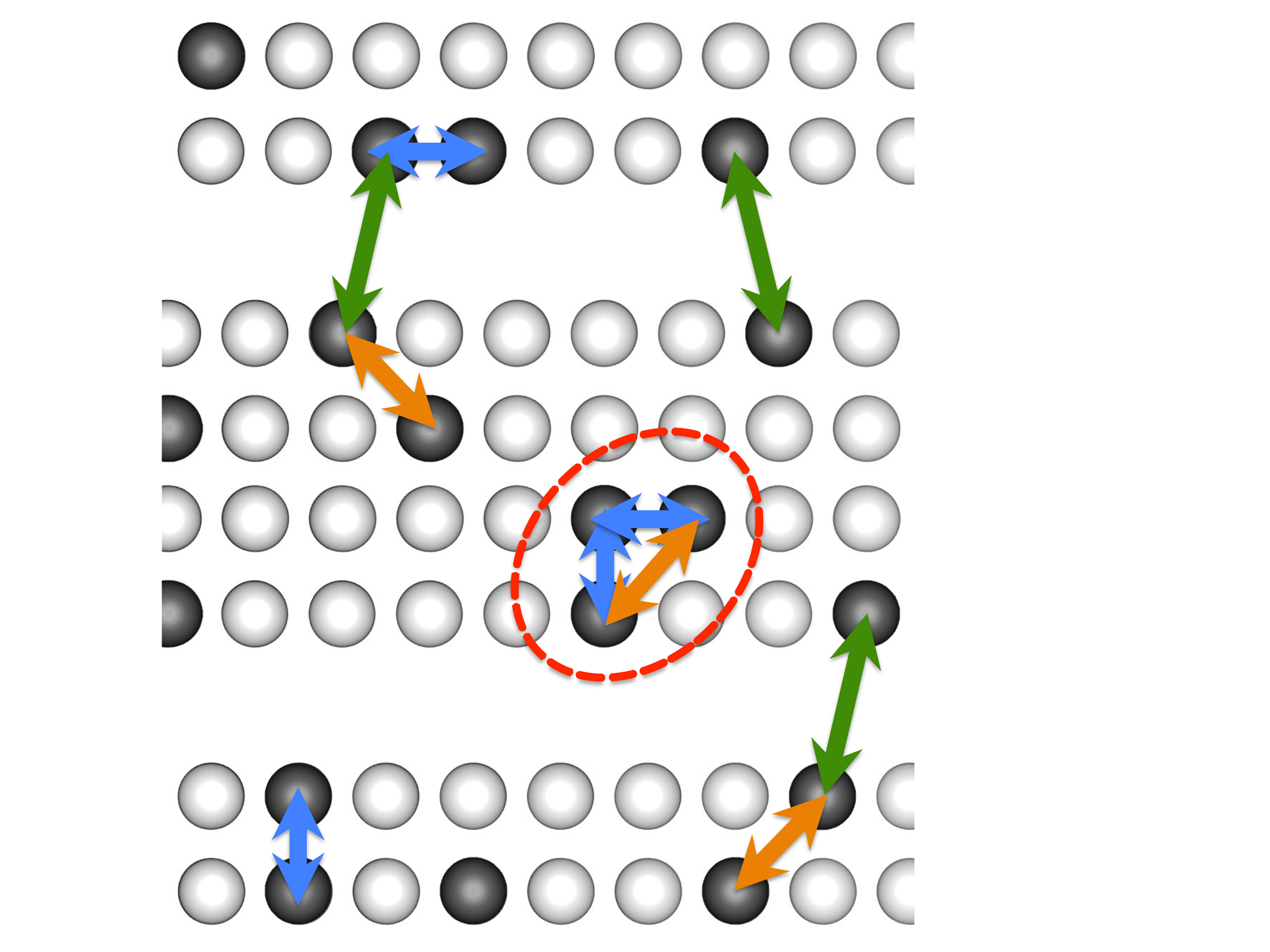}
\caption{(Color online) A sketch of the model for the network of $B$
  sites in the 4-layered Aurivillius structure considered in our Monte
  Carlo simulations. Black (white) spheres represent magnetic
  (non-magnetic) cations. The three types of coupling included in our
  model are indicated by double arrows: nearest-neighbor
  $J_{\mathrm{NN}}$ in blue, next-nearest-neighbor $J_{\mathrm{NNN}}$
  in orange, and inter-layer $J_{\mathrm{INTER}}$ in green. Not
  visible in the picture but included in the calculations are the
  $J_\text{NNN}$ couplings within the same in-plane perovskite
  layer. The red circle highlights a triangle of connected spins that
  can in principle give rise to partial frustration in the case of
  antiferromagnetic coupling.}
\label{fig:couplings}
\end{figure} 

Thus, the minimal model necessary to describe long-range magnetic
order in 4-layered Aurivillius phases with a concentration of $x=0.25$
magnetic cations on the $B$ sites must involve at least the following
three couplings: $J_{\text{NN}}$, $J_{\text{NNN}}$, and
$J_{\text{INTER}}$. Other further-neighbour couplings are either
weaker, or are not relevant for achieving long-range order, and are
therefore neglected within our model.
The three types of coupling that we consider in our model are
illustrated in Fig.~\ref{fig:couplings}.

The Hamiltonian of the system is then expressed in the standard
Heisenberg form:
\begin{equation}
\mathcal{H} = -\frac{1}{2}\sum_{i,j} J_{i,j} \mathbf{s}_{i} \cdot
\mathbf{s}_{j} - \Delta \sum_{i} (s^z_{i})^2 \quad ,
\label{Eq:Hamiltonian}
\end{equation}
where $\mathbf{s}_{i}$ is a three-dimensional unit-length vector
describing the direction of the magnetic moment at site $i$, and the
summation is over all $B$ sites occupied by magnetic ions. Here,
$J_{i,j}$ can have the values $J_{\text{NN}}$, $J_{\text{NNN}}$, or
$J_{\text{INTER}}$, depending on whether $i$ and $j$ are,
respectively, nearest-neighbors, next-nearest-neighbors within the
$m$-perovskite blocks, or closest possible neighbors across a
\bio-layer. In all the other cases, $J_{i,j}$ is zero. We also
consider a small easy axis anisotropy, $\Delta$, in
Eq.~\eqref{Eq:Hamiltonian}, in order to have a well-defined order
parameter along the $z$ direction.

Unless stated otherwise, the strengths of the three coupling constants
used in our simulations are set to values corresponding to those
obtained from {\it ab-initio} calculations for \btfo\ in
Ref.~\onlinecite{Birenbaum:2014bl}: $J_{\text{NN}}=45$\,meV,
$J_{\text{NNN}}=1.35$\,meV$\equiv 3\,\%\,J_{\text{NN}}$, and
$J_{\text{INTER}}=0.45$\,meV$\equiv
1\,\%\,J_{\text{NN}}$. Furthermore, we set $\Delta = 0.45$\,meV.
Note that all quantities in Eq.~\eqref{Eq:Hamiltonian} are defined
considering $\mathbf{s}_i$ as a unit vector, while the size of the
magnetic moments is absorbed in the coupling constants $J_{i,j}$ and
in $\Delta$.

We note that the coupling constants obtained in
Ref.~\onlinecite{Birenbaum:2014bl} are all antiferromagnetic, i.e.,
corresponding to a negative sign of $J_{i,j}$ in
Eq.~\eqref{Eq:Hamiltonian}. This can result in closed loops of bonds
that are partially frustrated, e.g. bonds relative to the couplings
$J_{\text{NN}}$ and $J_{\text{NNN}}$ as shown for the bonds marked by
the red circle in Fig.~\ref{fig:couplings}. In order to avoid
complications in the Monte Carlo simulations due to such partial
frustration, and since we mostly want to provide an upper limit for
the transition temperature, we neglect the possible role of
frustration in decreasing $T_C$ by assuming all couplings to be
ferromagnetic, i.e. with positive sign.

\subsection{Monte Carlo Simulations}
 \label{subsec:monte_carlo_deets}

To obtain temperature dependent properties for the model described in
the previous subsection, we perform Monte Carlo simulations using the
Metropolis algorithm and parallel tempering.~\cite{Earl2005}
Furthermore, we average the so-obtained macroscopic quantities over
several realizations of the magnetic cation distribution. In practice,
we apply the following procedure:
\begin{enumerate}
\item Generate a random distribution of magnetic cations within our
  simulation cell (respecting the constraint discussed in the previous
  section).
\item Determine the network of bonds for each type of coupling.
\item Use Monte Carlo to calculate thermodynamic quantities for
  this specific configuration.
\item Repeat this procedure several times, each time
  generating a different random distribution of magnetic cations over
  the $B$ site positions.
\item Take the average of the quantities obtained for each
  specific cation distribution.
\end{enumerate}

Furthermore, we repeat this procedure for different system sizes to
obtain an accurate estimate of $T_{\text{C}}$ in the limit of infinite
system size. For this purpose we note that, due to the small easy-axis
anisotropy included into the model, the relevant order parameter is
$M^z =\frac{1}{N} \langle \langle \vert \sum_{i} s^z_{i} \rvert
\rangle \rangle_{\text{C}}$, the component of the magnetization along
the easy axis. Here, $\langle \dots \rangle$ indicates an average over
Monte Carlo measurements, while $\langle \dots \rangle_{\text{C}}$
indicates the average over different configurations, i.e., different
realizations of the magnetic cation distribution. We average over
different cation distributions to account for the self-averaging
present in macroscopically large samples.

Thus, $T_\text{C}$ can be determined from the crossing point of Binder
cumulants \cite{Binder1981,Binder1981PRL} of $M^z$ calculated for
different system sizes,\cite{Binder2010} where the Binder cumulant is
given by:
\begin{equation}
B_C = \left< 1 - \frac{\left< \left(\sum_{i} s^z_{i} \right)^4
  \right>} {3 \left< \left( \sum_{i} s^z_{i} \right)^2 \right>^2
}\right>_{\text{C}} \quad .
\label{Eq:BinderC}
\end{equation}
Furthermore, we also calculate the magnetic susceptibility:
\begin{equation}\label{Eq:mag_sus}
\chi = \frac{1}{N k_B T} \left< \left< \left(\sum_i \mathbf{s}_i
\right)^2 \right> - \left< \lvert \sum_i \mathbf{s}_i \rvert \right>^2
\right>_{\text{C}} \quad ,
\end{equation}
where $T$ is the temperature and $k_B$ is the Boltzmann constant.

Typically, $80$ temperatures are run in parallel using the parallel
tempering procedure. Temperatures are distributed exponentially
according to $T_l = T_{\text{min}} \alpha^{(l-1)}$, where
$l=1,\dots,80$ and $\alpha > 1$.
We use the same values for $T_{\text{min}}$ and $\alpha$ for all
simulations with the same $x$, independent of the size of the
system. Measurements of magnetization and energy are collected every
$300$ sweeps (average number of update trials per spin). A certain
number of initial measurements (typically $4\cdot10^3$) are not
included in the averages for thermalization purposes, and averages are
taken over a number of measurements varying between $10^4$ and
$1.9\cdot 10^5$, depending on system size.

\section{Results and Discussion}

\subsection{Transition temperature for \btfo}
\label{sec:BTFO}

\begin{figure}
\centering
\includegraphics[width=1\columnwidth]{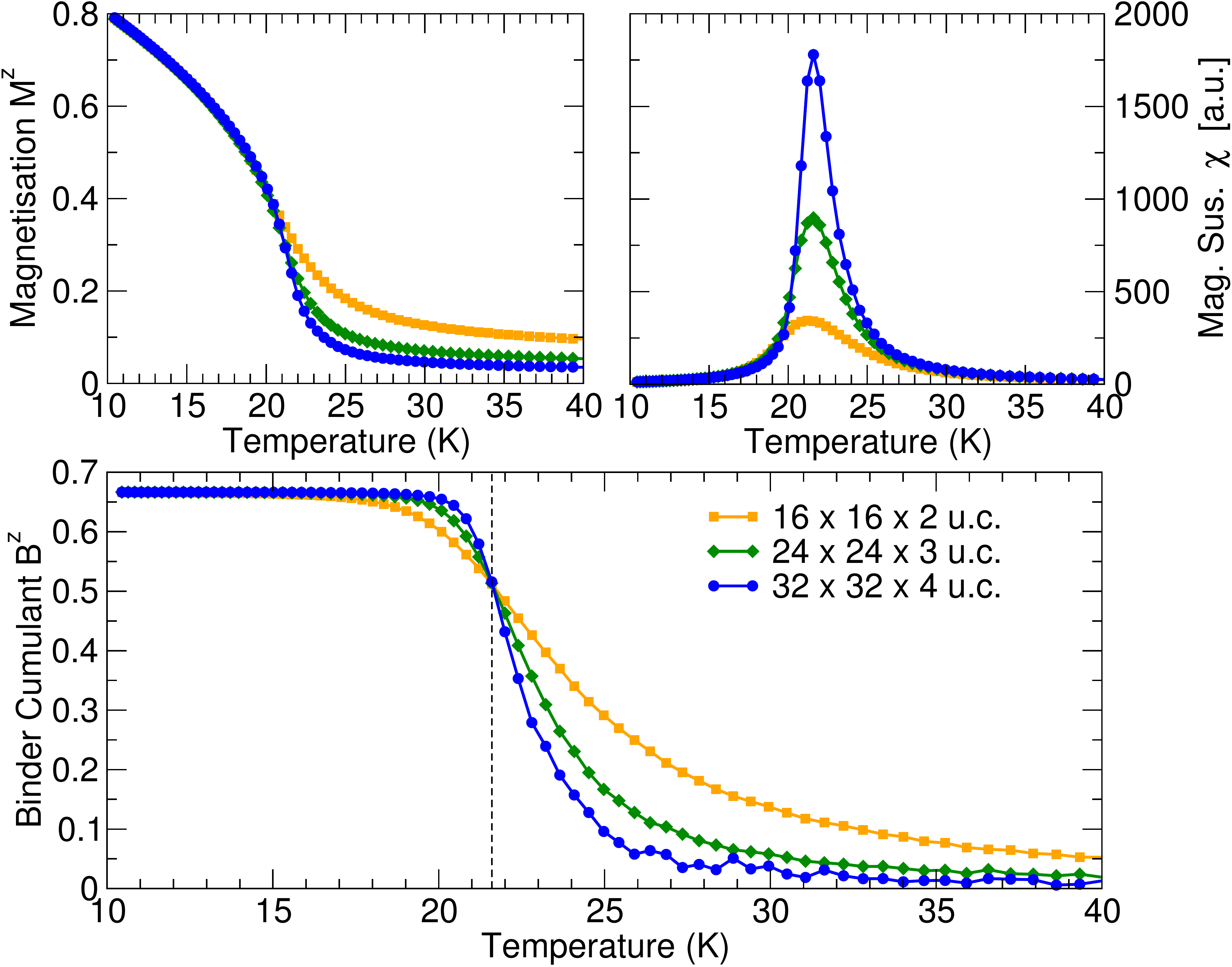}
\caption{(Color online) Temperature dependence of a) magnetization, b)
  magnetic susceptibility and c) the Binder cumulant for the model
  described in Sec.~\ref{sec:theo_meth} and magnetic ion concentration
  $x=0.25$. Several system sizes have been considered: $8n \times 8n
  \times n$ unit cells with $n=2$ ($\blacksquare$) in orange, $n=3$
  ($\blacklozenge$) in green, and $n=4$ ($\bullet$) in blue. The
  vertical dashed line in c) indicates $T_\text{C}$, obtained from the
  intersection point of the Binder cumulants for the three different
  cell sizes.}
\label{fig:BC_BFTO}
\end{figure}

We begin the discussion of our results by considering the case of
\btfo, i.e. a concentration of $x=0.25$ and all coupling constants
fixed to the approximate values calculated in
Ref.~\onlinecite{Birenbaum:2014bl}, as specified in
Sec.~\ref{subsec:parameters}. Fig.~\ref{fig:BC_BFTO} shows the
temperature dependence of the magnetization, magnetic
susceptibilities, and Binder cumulants for different sizes of the
simulation cell. In each case, all quantities are obtained by
averaging over an ensemble of 5 configuration corresponding to
different distributions of the \fe\ cations.

From the crossing point of the Binder cumulants for the three
different system sizes, we obtain a magnetic transition temperature
$T_\text{C}=22$\,K. This value is also consistent with the temperature
dependence of the magnetization and with the peak positions of the
magnetic susceptibility, which appear to depend only weakly on system
size.

The value of $22$\,K is more than two orders of magnitude smaller than
$J_{\text{NN}}$. This small value of the transition temperature arises
from the fact that the concentration of Fe$^{3+}$ in \btfo\ is too low
to allow percolation of the nearest-neighbor bonds within the
$m$-perovskite blocks (see the discussion in the previous section).
Therefore, the system essentially consists of small isolated clusters
of \fe, where the magnetic moments within each cluster are coupled
strongly through $J_{\text{NN}}$, but different clusters within the
same $m$-perovskite block are only weakly coupled through
$J_{\text{NNN}}$. Furthermore, different 4-perovskite blocks are
coupled through the very weak interaction $J_{\text{INTER}}$.

This low value of $T_C$, which, based on the nature of our
approximations, can be viewed as an upper limit to the magnetic
transition temperature of the real material, indicates that claims of
room-temperature magnetism measured in \btfo\ are very unlikely
related to the intrinsic properties of this material. Instead,
inclusions and other magnetic impurities are most likely responsible
for the apparent high temperature magnetic properties. We also note
that the superexchange interaction between \fe\ cations is generally
one of the strongest among all $3d$ transition metal cations, and thus
the same conclusion holds for similar reports on related 4-layered
Aurivillius phases containing other $M^{3+}$ $3d$ transition metal
cations with concentration $x=0.25$.

In the following, we establish the sensitivity of the calculated
$T_\text{C}$ on the specific values used for $J_{\text{NNN}}$ and
$J_{\text{INTER}}$. This also allows to identify which coupling
constitutes the more severe bottleneck for achieving higher transition
temperatures and thus should possibly be increased, e.g. by applying
strain or pressure, to best engineer 4-layered Aurivillius phases with
magnetic order at higher temperatures.

\subsection{Dependence of $T_\text{C}$ on $J_{\text{NNN}}$ and $J_{\text{INTER}}$}
\label{ssec:TNdependence}

Next, we investigate the dependence of the transition temperature on
the weak further-neighbor couplings $J_{\text{NNN}}$ and
$J_{\mathrm{INTER}}$. To this end, we perform analogous simulations as
described on the previous section, but with varying strength for one
of these coupling constants, while the other one is set to be equal to
$J_\text{NN}$. In this way, there is always only one ``weak'' coupling
in the system, and one can observe how $T_\text{C}$ is reduced on
decreasing the strength of this particular coupling constant.
These calculations are performed for a system containing $N = 24
\times 24 \times 3 $ basic unit cells, and concentration of magnetic
ions fixed to $x=0.25$. Since the purpose of these calculations is not
to obtain an extremely accurate value for $T_\text{C}$, but rather to
observe the trend as the coupling constants are tuned, we do not
obtain the transition temperature using the intersection of Binder
cumulants for different system sizes. Instead, we extract the peak
position of the magnetic susceptibility for eight different random
distributions of the magnetic ions. We then estimate the transition
temperature by averaging this peak position over all eight
configurations, and we estimate the corresponding error using the
standard deviation of the different values.

\begin{figure}
\centering
\includegraphics[width=\columnwidth]{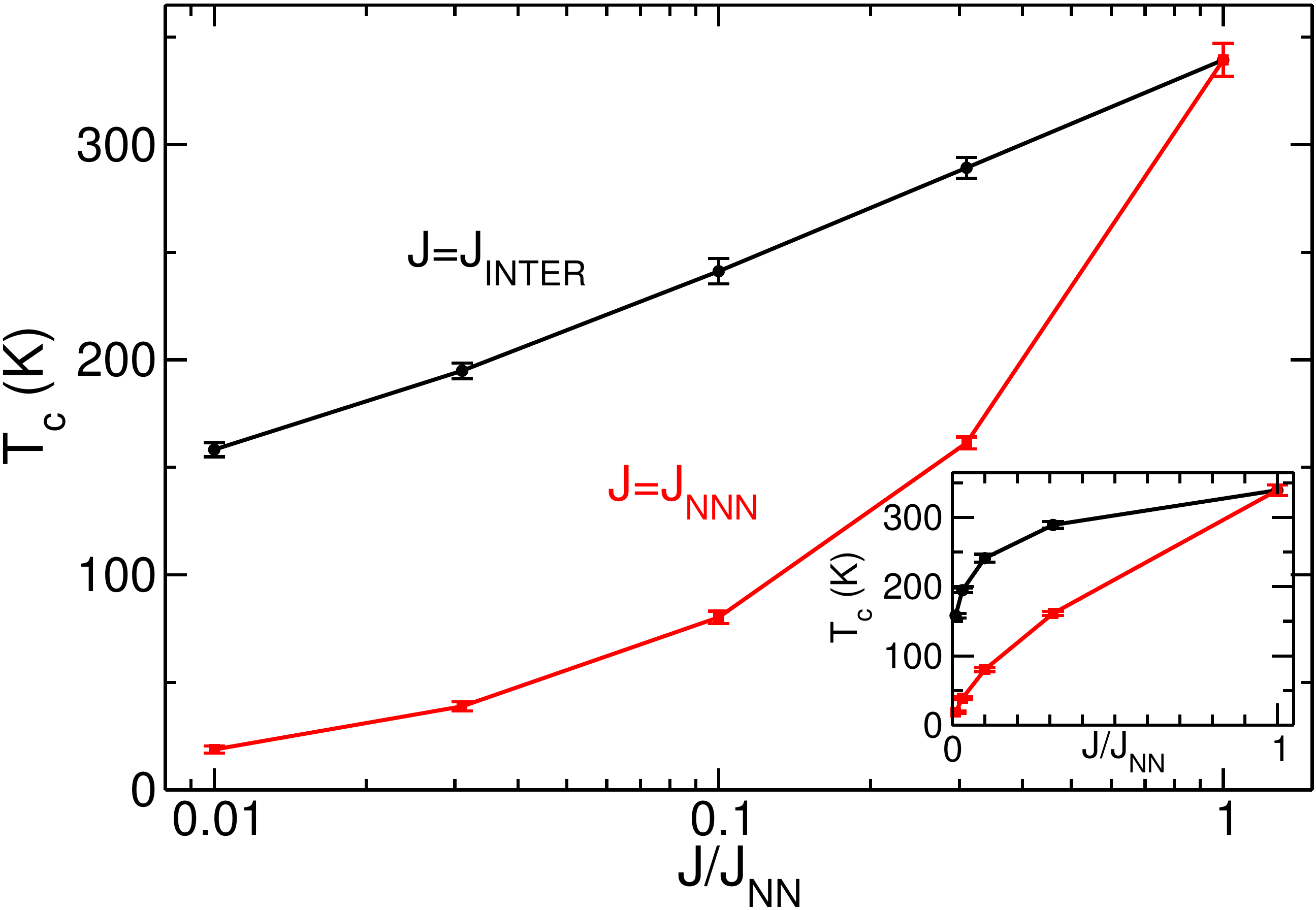}
\caption{(Color online) Dependence of the transition temperature,
  $T_\text{C}$, on $J_{\text{NNN}}$ ($J_{\text{INTER}}$) for
  $J_{\text{INTER}}=J_{\text{NN}}=45$ meV
  ($J_{\text{NNN}}=J_{\text{NN}}= 45$ meV). The red squares (black
  dots) indicate the calculated values whereas the black (red) line is
  just a guide for the eyes. Error bars are estimated from the
  standard deviation of the peak positions of the magnetic
  susceptibility obtained for several different Fe$^{3+}$
  distributions. The inset shows the same data on a linear
  scale.}
\label{fig:NNNvsINTER}
\end{figure}

We first consider the case where all couplings have the same size as
$J_{\text{NN}}$,
i.e. $J_{\mathrm{NNN}}=J_{\mathrm{INTER}}=J_{\text{NN}}=45$ meV.  It
can be seen from Fig.~\ref{fig:NNNvsINTER} that, although the overall
energy scale of the couplings is quite large ($\sim$520\,K), we obtain
a transition temperature $T_\text{C}\approx 340$ K, which is only
slightly above room temperature. This is due to the low concentration
of magnetic ions, resulting in a low average coordination number of
the magnetic lattice.

Keeping $J_{\mathrm{NN}} = J_{\mathrm{NNN}} = 45$ meV, we then
decrease $J_{\mathrm{INTER}}$ in several steps from 45\,meV to
0.45\,meV. The resulting transition temperatures are shown as black
dots in Fig.~\ref{fig:NNNvsINTER}.  Interestingly, the transition
temperature exhibits only a moderate decrease (by approximately a
factor 0.5) although $J_{\text{INTER}}$ is decreased by two orders of
magnitude.

Next, we keep $J_{\mathrm{INTER}} = J_{\mathrm{NN}} = 45$\,meV and
decrease the value of $J_{\text{NNN}}$ stepwise from 45\,meV to
0.45\,meV. The resulting transition temperatures are indicated as red
squares in Fig.~\ref{fig:NNNvsINTER}. We observe that, in this case,
the value of $T_\text{C}$ decreases more dramatically compared to the
previous case where $J_{\text{INTER}}$ is decreased.

The reason for this profound difference regarding the sensitivity of
$T_\text{C}$ on $J_{\text{NNN}}$ compared to $J_{\mathrm{INTER}}$ can
be qualitatively explained as follows. For $J_{\mathrm{INTER}}=0$, the
system consists of uncoupled 4-layers, while the network of bonds
created by $J_{\text{NN}}$ and $J_{\text{NNN}}$ percolate each
individual 4-layer. Thus, the system essentially represents a
quasi-two-dimensional system (see
Refs.~\onlinecite{Yasuda:2005dw,Wehinger2016}). In this case, and
considering isotropic spins (i.e., neglecting the small easy axis
anisotropy in the model), the correlation length perpendicular to $c$
would diverge as $T\rightarrow 0$. Thus, when $J_{\mathrm{INTER}}$ is
switched on, the effective net interaction between the 4-layers, and
thus the energy-scale determining the temperature for long-range
magnetic ordering, is not simply given by $J_\text{INTER}$. Instead,
it is given by $J_{\text{INTER}}$ multiplied by the number of
correlated spins within each layer, i.e. all spins within a radius of
the size of the correlation length at that temperature. This implies
that the transition temperature has a relatively weak dependence on
the inter-layer coupling.

A similar scenario does not occur, however, in the limit of small
$J_{\text{NNN}}$. In this case, the bonds created by $J_{\text{NN}}$
and $J_{\text{INTER}}$ are not sufficient to overcome the percolation
threshold within the system, both within the 4-layers and along
$c$. Thus, for $J_{\text{NNN}}=0$, the system consists of a
conglomeration of uncoupled clusters, and the correlation length is
limited by the average size of these clusters.
For $J_\text{NNN} \neq 0$ and temperatures at which the correlation
length is of the size of each cluster, the system resembles isolated
magnetic moments of different sizes (corresponding to the cluster
sizes) coupled by $J_{\text{NNN}}$. This would imply a linear
dependence of $T_\text{C}$ on $J_{\text{NNN}}$. However, since the
concentration of $x=0.25$ is relatively close to $x_c \sim 0.311$, the
typical size of the clusters might be relatively large and thus the
temperature at which all spins within the clusters are fully
correlated might be quite small.

\begin{figure}
\centering
\includegraphics[width=\columnwidth]{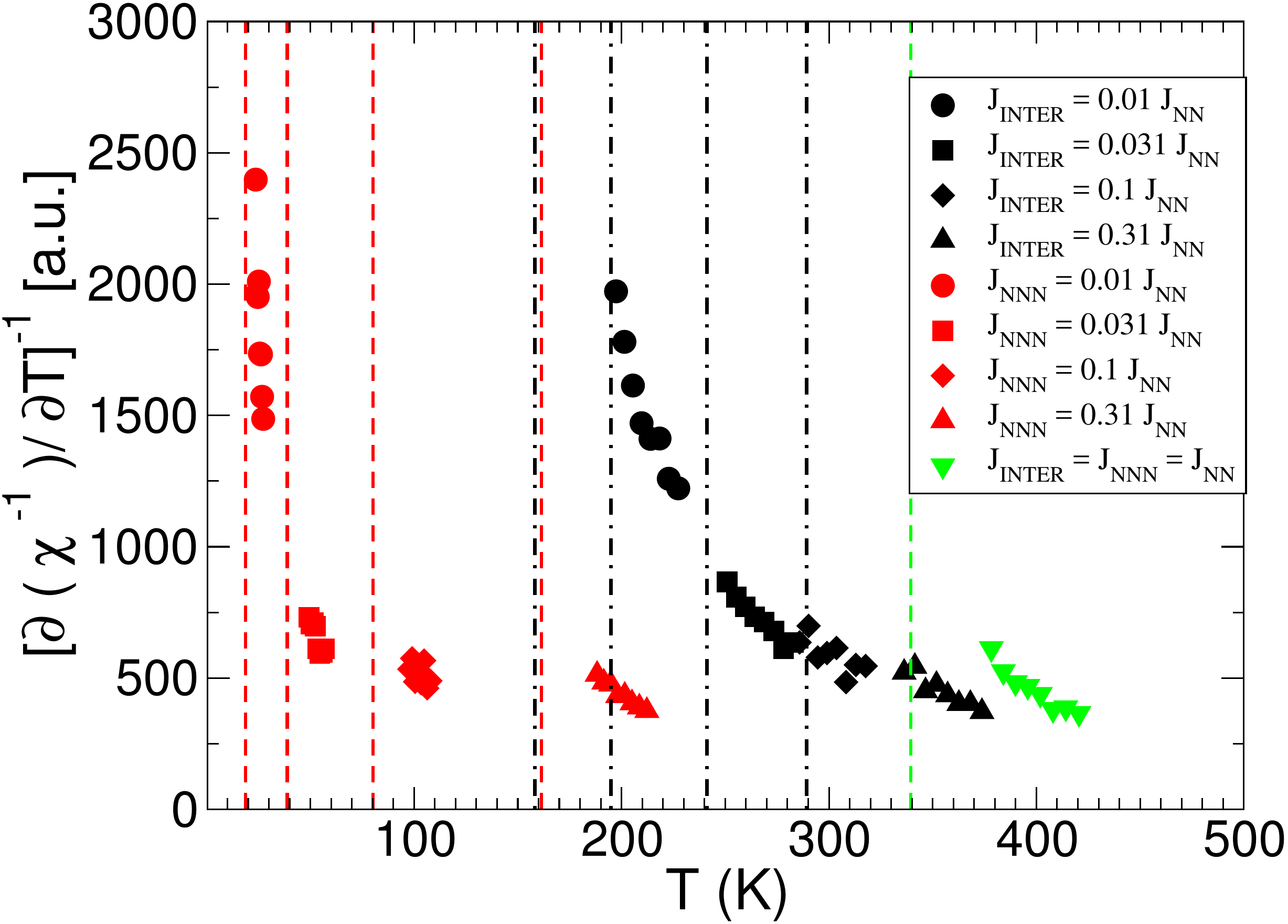}
\caption{(Color online) Values of the numerical derivative $[ \partial
    \chi^{-1} / \partial T ]^{-1} $ at different temperatures for the
  cases where either $J_{\text{INTER}}$ (black points) or
  $J_{\text{NNN}}$ (red points) is varied. The vertical lines indicate
  the transition temperatures for the various cases as shown in
  Fig.~\ref{fig:NNNvsINTER}}
\label{fig:InvDerInv}
\end{figure}

This scenario is qualitatively supported by the behavior of the
magnetic susceptibility above $T_\text{C}$, calculated for different
values of $J_\text{NNN}$ and $J_\text{INTER}$. Within this
paramagnetic regime, the susceptibility is expected to exhibit
Curie-Weiss behavior, i.e., $\chi(T)=\frac{C}{T-T_\text{C}}$, where
the constant $C$ is proportional to the square of the effective
magnetic moments. Thus, if interactions occur between clusters of
correlated spins, one expects $C$ to be quite large.
We can extract $C$ by taking the inverse of the numerical derivative
of $\chi^{-1}(T)$, i.e. $[ \partial \chi^{-1} / \partial T
]^{-1}$. Fig.~\ref{fig:InvDerInv} shows the so-obtained values for
temperatures above $T_\text{C}$ and for the various values of
$J_{\text{NNN}}$ and $J_{\text{INTER}}$. It can be seen that one
obtains rather large ``effective moments'' when $J_\text{INTER}$ is
decreased, even at temperatures around 200\,K. In contrast, the
effective moments remain small down to rather low temperatures on
decreasing $J_\text{NNN}$.
This indicates that, at a given temperature, the size of the
correlated clusters of spins is much larger when the system resembles
weakly coupled two dimensional layers, than when the system resembles
weakly coupled non-percolating clusters.

We point out that the realistic values for $J_{\text{NNN}}$ and
$J_{\text{INTER}}$ ($1.35$\,meV and $0.45$\,meV, respectively) are
both very low compared to the strong nearest-neighbor coupling
$J_{\text{NN}}$ ($45$\,meV) in \btfo.  Therefore, the real system
corresponds to weakly coupled layers formed by weakly coupled
clusters, which explains the low transition temperature found in
Sec.\ref{sec:BTFO}. Furthermore, this suggests that a large
enhancement of $J_{\text{INTER}}$ and in particular of
$J_{\text{NNN}}$ would be necessary to increase the transition
temperature towards higher values. While it is in principle
conceivable to enhance $J_\text{NNN}$ and $J_\text{INTER}$ via
pressure, strain, doping, or cation substitution, it appears quite
unlikely that the required order-of-magnitude changes can be achieved
in this way. Therefore, the most promising route to obtain Aurivillius
phases with magnetic long-range order at or above room temperature is
to increase the concentration of magnetic ions.

\subsection{Aurivillius phases with higher concentration of magnetic cations}
\label{sec:higher_conc}

\begin{figure*}
\centering
\includegraphics[width=\textwidth]{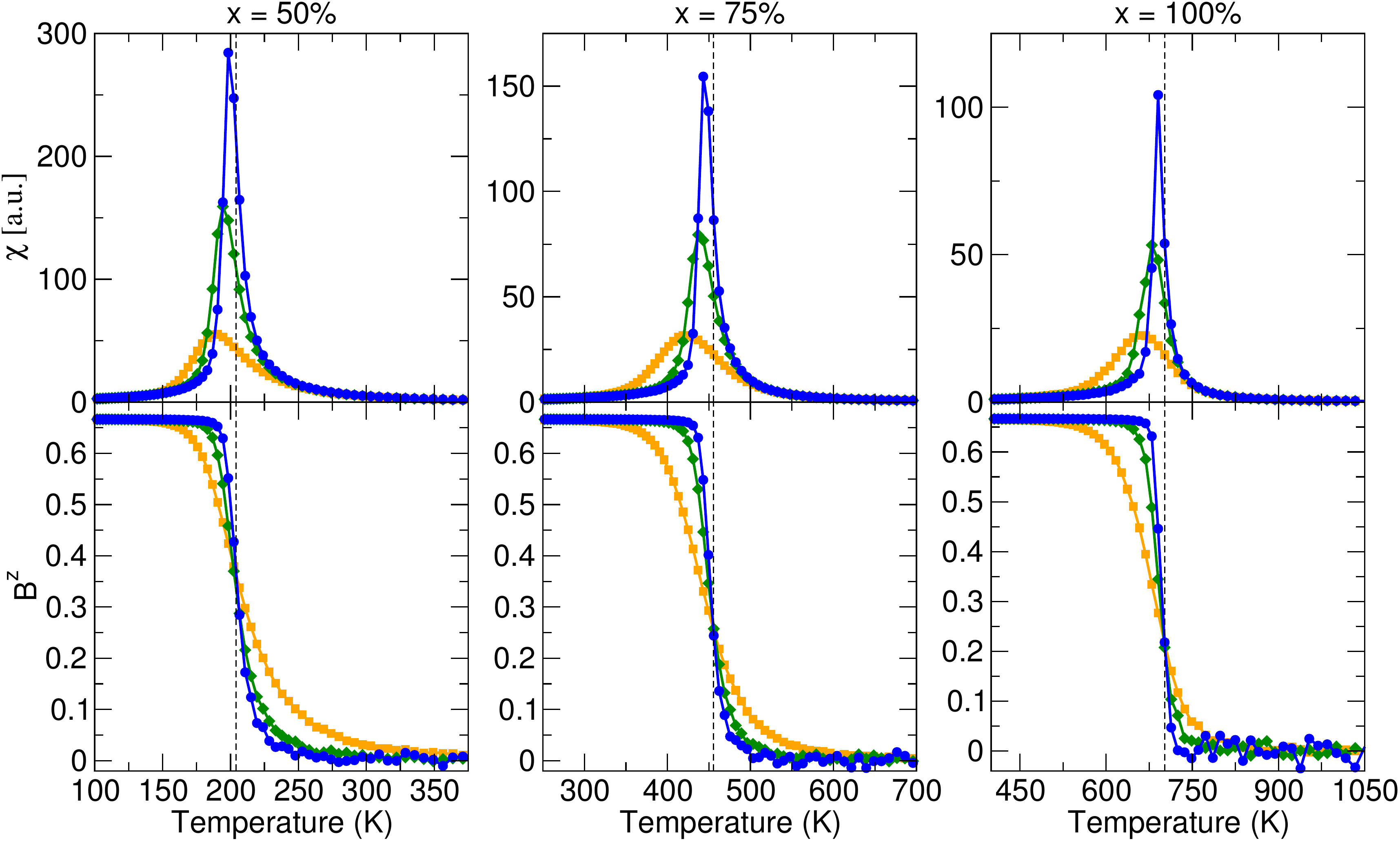}
\caption{(Color online) Temperature dependence of the magnetic
  susceptibility, Eq.~\eqref{Eq:mag_sus}, (top row) and Binder
  cumulant, Eq.~\eqref{Eq:BinderC}, (bottom row) for magnetic ion
  concentrations $x=0.5$, $0.75$, and $1.0$ in the $m=4$ Aurivillius
  structures. Different system sizes have been considered: $8n \times
  8n \times n$ unit cells with $n=2$ in orange ($\blacksquare$), $n=3$
  in green ($\blacklozenge$), and $n=4$ in blue ($\bullet$). Vertical
  lines represent the obtained $T_\text{C}$.}
\label{fig:t_c-concentration}
\end{figure*}
	
Motivated by the results presented in the previous sections, we now
consider also magnetic ion concentrations larger than 25\,\%.
Generally, the achievable concentration of magnetic cations in
Aurivillius phases is constrained by stoichiometry and size
restrictions.
Only cations within a certain size range have been found suitable to
occupy the $B$ sites within the Aurivillius
structure\cite{Newnham/Wolfe/Dorrian:1971}. Most magnetic cations that
are within this suitable range have a valence of $+3$, e.g. Fe$^{3+}$,
Co$^{3+}$, Cr$^{3+}$, or Mn$^{3+}$.
However, a total of $6m$ electronic charges have to be balanced by the
cations occupying the $(m-1)$ $A$ sites and the $m$ $B$ sites. Typical
$A$ site cations in the Aurivillius phases have either a valence of
$+3$ or $+2$, with Bi$^{3+}$ perhaps being most common. Thus, with
Bi$^{3+}$ (or any other possible $3+$ cation) on the $A$ site, the
required average $B$ site valence is given by $q_B = 3(m+1)/m$.
This implies that it is not possible to have all available $B$ sites
occupied with magnetic $3+$ cations (except in the limit $m
\rightarrow \infty$). Instead, a certain percentage of $B$ sites has
to be occupied by nonmagnetic cations with higher valence, such as
e.g. \ti\ as in \btfo. Note that it has been found impossible to
incorporate significant amounts of the higher-valent (and rather
small) Mn$^{4+}$ cation into the Aurivillius structure
\cite{McCabe/Greaves:2005,Zurbuchen_et_al:2007}.

Based on these restrictions, there are in principle two ways to
increase the concentration of magnetic cations. The first possibility
is to increase $m$, the number of layers within the perovskite blocks,
which decreases the required average $B$ site valence $q_B$. Indeed,
magnetic long-range order has been reported for several $m=5$
systems.\cite{Jartych_et_al:2013,Keeney_et_al:2013} The second
possibility, which we explore in the remainder of this article, is to
keep $m$ fixed and substitute \ti\ with nonmagnetic cations of even
higher valence, such as Nb$^{5+}$, Ta$^{5+}$, Mo$^{6+}$, or W$^{6+}$.
For $m=4$ this leads to compositions such as
e.g. Bi$_5$Fe$_{1+x}$Ti$_{3-2x}$Nb$_x$O$_{15}$ or
Bi$_5$Fe$_{1+2x}$Ti$_{3-3x}$W$_x$O$_{15}$ (\mbox{$0 \leqslant x
  \leqslant 1$}). In Table~\ref{table:concentration}, we propose some
examples for Aurivillius phases that correspond to higher magnetic
cation concentration and higher critical temperatures. While (with
\btfo 's exception and to the best of our knowledge) none of these
compounds have been synthesized yet, the suggested cations have all
been successfully incorporated on the $B$ sites of other known
Aurivillius phases.

\begin{table*}
\centering
\caption{\label{table:concentration} Proposed examples of Aurivillius
  phases with different magnetic ion concentrations, and their
  expected transition temperatures, $T_C$. While (except for the case
  of \btfo and to the best of our knowledge) these compositions have
  not been synthesized, yet, all of the corresponding $B$ site cations
  have been successfully incorporated in other Aurivillius compounds.}
\begin{ruledtabular}
\begin{tabular}{rcccc}
	\toprule		
	estimated $T_C$ (K)   & 22    & 204   & 455   & 701 \\
	concentration (\%)  & 25    & 50    & 75    & 100 \\
	example Aurivillius & \btfo & $\mathrm{Bi_5 Fe_2 Nb Ti O_{15}}$ & $\mathrm{Bi_5 Fe_3 W O_{15}}$ & ?  \\
	\bottomrule
\end{tabular}
\end{ruledtabular}
\end{table*}	

In the following we assume that increasing the number of \fe\ (or
other magnetic cations) and replacing part of the \ti\ with
higher-valent non-magnetic cations will not significantly alter the
magnitude of the three relevant magnetic coupling constants considered
in our model. We thus keep the values for $J_{\text{NN}}$,
$J_{\text{NNN}}$ and $J_{\text{INTER}}$ fixed to the ones derived from
the {\it ab-initio} calculations for \btfo.~\cite{Birenbaum:2014bl}
The most relevant effect resulting from the increased magnetic ion
concentration that is included in our model is therefore the
increasing amount of bonds between magnetic ions coupled through the
strong nearest-neighbor coupling $J_{\text{NN}}$.

We consider the concentrations $x=0.5$, $x=0.75$, and $x=1$, and
perform Monte Carlo simulations for supercells of sizes
$16\times16\times2$, $24 \times 24 \times 3$, and $32 \times 32 \times
4$ in units of the basic cell. Similar to what was discussed in
Sec.~\ref{sec:BTFO}, the macroscopic averages obtained for $x=0.5$ and
$x=0.75$ are averaged over five different random distributions of
magnetic cations.
 
Fig.~\ref{fig:t_c-concentration} shows the temperature dependence of
the magnetic susceptibilities (upper panels) and Binder cumulants
(lower panels) for the three concentrations of magnetic ions on the
$B$ sites. For all three concentrations, the Binder cumulants obtained
for differently sized simulation cells intersect at a temperature very
close to the peak position of the magnetic susceptibility calculated
for the largest system size. This indicates that finite size effects
are relatively small in this case.
Furthermore, the transition temperature for $x=0.5$, i.e., when half
of all $B$ sites are occupied by magnetic ions, is significantly
higher than for $x=0.25$, but still below room temperature. However,
for $x=0.75$, long-range magnetic order appears at temperatures well
above 100$^\circ$C. Finally, when all $B$ sites are occupied by
magnetic ions, i.e., for $x=1.0$, we obtain $T_\text{C}=701$\,K.

\begin{figure}
\centering
\includegraphics[width=\columnwidth]{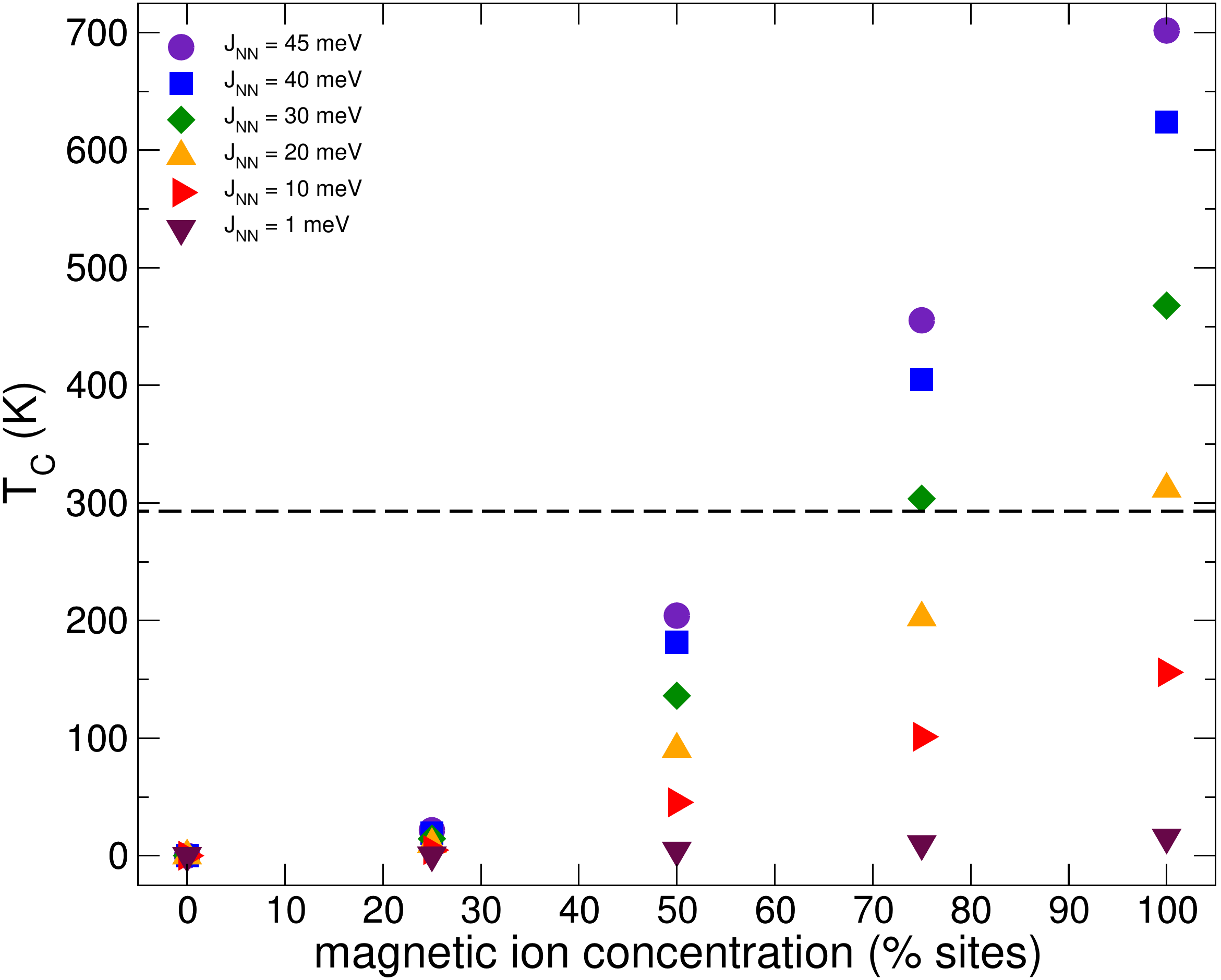}
\caption{(Color online) Magnetic transition temperature, $T_\text{C}$,
  as function of magnetic ion concentration. Different symbols specify
  temperatures for different (scaled) values of $J_{\mathrm{NN}}$
  (with $J_{\mathrm{NNN}} = 3\% \, J_{\mathrm{NN}}$ and
  $J_{\mathrm{INTER}} = 1\% \, J_{\mathrm{NN}}$). The dashed horizontal
  line corresponds to room temperature.}
\label{fig:t_c-concentration-couplings}
\end{figure}

The transition temperatures obtained for different concentrations are
summarized in Fig.~\ref{fig:t_c-concentration-couplings}. Since the
energy scale (and thus the specific value of $T_\text{C}$) in the
Monte Carlo simulations is defined via $J_\text{NN}$, we also include
scaled values of $T_\text{C}$, which correspond to different strength
of $J_\text{NN}$ (but fixed ratios $J_\text{NNN}/J_\text{NN}$ and
$J_\text{INTER}/J_\text{NN}$), e.g. for magnetic $M^{3+}$ cations
other than \fe.
We note that, for concentrations above the critical value for
percolation of the nearest-neighbor bonds ($x_c \approx 0.312$), the
transition temperature seems to increase linearly with concentration.

Measurements of magnetic properties on Aurivillius phases
Bi$_{m+1}$Ti$_{3}$Fe$_{m-3}$O$_{3m+3}$ performed by Jartych et
al. \cite{Jartych_et_al:2013} indicate transition temperatures to a
spin glass state of $T_\text{N}=260$\,K and $T_\text{N}=280$\,K for
$m=6$ and $m=7$ ($x=0.5$ and $x=0.571$), respectively. These values
are quite comparable in size to the transition temperature we obtain
for $x=0.5$ (see Fig.~\ref{fig:t_c-concentration-couplings} and
Table~\ref{table:concentration}).
Furthermore, the value of $T_\text{C}=701$\,K obtained for $x=1.0$ is
rather similar to the N{\'e}el temperature of the $m \rightarrow
\infty$ perovskite BiFeO$_3$ ($643$ K) \cite{Roginskaya1966}. 

Even though this comparison should be taken with care, since the
values obtained from our simulations should merely be interpreted as
upper bounds for the transition temperatures of the real materials
(with partially frustrated antiferromagnetic interactions), this seems
to indicate the following: 1) the magnitudes we used for
$J_{\text{NN}}$ and the assumption that $J_{\text{NN}}$ depends only
weakly on concentration are indeed reasonable, 2) the weak inter-layer
coupling is not prohibitive for achieving transition temperatures
around or above room temperature, and 3) the partial frustration of
antiferromagnetic bonds is not very strong, at least for cases with
$x\sim0.5$.

\section{Summary and Conclusion} 
\label{sec:conclu}

In this work, we have, from a theoretical perspective, explored
whether Aurivillius phases can exhibit magnetoelectric multiferroic
states at or above room temperature. To this end, we have established
a Heisenberg model corresponding to magnetically-dilute 4-layered
ferroelectric Aurivillius phases.
The minimal model for which magnetic long-range order can occur within
these compounds, including the important case with only 25\,\% of all
$B$ sites occupied by magnetic cations, requires the presence of
nearest-neighbor, next-nearest-neighbor, and inter-layer couplings.
To obtain the corresponding magnetic transition temperatures, we have
performed Monte Carlo simulations, thereby averaging over several
distributions of magnetic cations over the available $B$ sites. We
obtain upper limits for the transition temperature by neglecting the
partial frustration that can occur in the case of
antiferromagnetically coupled spins.

For the case of \btfo (concentration $x=0.25$) we use coupling
constants based on earlier {\it ab-initio} calculations, and we obtain
a transition temperature of $T_\text{C}=22$\,K, i.e. far below room
temperature. In order to identify the most promising strategy for
achieving magnetic long-range order at higher temperatures, we have
then addressed the individual effects of the weak
next-nearest-neighbor coupling within the 4-perovskite blocks,
$J_\text{NNN}$, and of the weak inter-layer coupling between these
blocks, $J_\text{INTER}$.
Our results indicate that the most crucial coupling in the dilute
($x=0.25$) case is $J_{\rm NNN}$. Even though the presence of the
inter-layer coupling is crucial to achieve percolation along the $c$
direction, the strength of $J_\text{NNN}$ has a much stronger impact
on the transition temperature. A significant increase of
$J_{\text{NNN}}$ with respect to the value obtained from {\it
  ab-initio} calculations for \btfo\ seems necessary to achieve
magnetic order around room temperature. However, it is unclear whether
and how such a significant increase of $J_\text{NNN}$ could be
realized.

Therefore, we have explored a more promising route toward higher
$T_\text{C}$, which is to increase the concentration of magnetic ions
within the $m$-perovskite blocks. For the $m=4$ case considered here,
our results indicate that for $x \gtrsim 0.6$, magnetic transition
temperatures around or above room temperature can be reached.
To obtain 4-layered Aurivillius phases with increased magnetic ion
concentrations, we suggest to combine trivalent magnetic $3d$
transition metal cations such as \fe\ with high-valent non-magnetic
cations such as e.g. Nb$^{5+}$, Ta$^{5+}$, Mo$^{6+}$, or W$^{6+}$. The
calculated transition temperatures as well as some suggested
compositions with varying magnetic ion concentrations are listed in
Table~\ref{table:concentration}. It can be seen that for Aurivillius
phases with $x=0.75$ (e.g. Bi$_5$Fe$_3$WO$_{15}$) magnetic transition
temperatures well above room temperature can be expected.

Finally, our results demonstrate that the weak inter-layer coupling
between adjacent $m$-perovskite blocks is not prohibitive for
achieving long-range magnetic order above room temperature. This is
consistent with previous studies of quasi-two-dimensional Heisenberg
systems.~\cite{Yasuda:2005dw,Wehinger2016}
Here, we extend these studies to the case of a quasi-two-dimensional
dilute magnetic system, albeit with the simplification of using only
ferromagnetic interactions and thus excluding the case with partially
frustrated interactions. We hope that our work will stimulate further
research in this interesting direction.
We also note that, even though we considered only the specific example
of a 4-layered Aurivillius structure, our results allow for some
generalization to other $m$ values and also to other families of
layered perovkite systems, such as the Ruddlesden-Popper or
Dion-Jacobson series.~\cite{Benedek_et_al:2015} These are two other
examples of naturally-layered oxides consisting of a certain number of
perovskite layers, stacked along [001], and separated by different
inter-leaving layers. In particular, the connectivity between the
octahedrally coordinated cation sites in the Ruddlesden-Popper series
is equivalent to the present case of the Aurivillius structure, and
thus our minimal Heisenberg model is also applicable to these systems.

\begin{acknowledgments}
This research was supported by ETH Zurich and the Swiss National
Science Foundation through Grant No. 200021\_141357 and through
NCCR-MARVEL.
\end{acknowledgments}

\bibliography{references}

\end{document}